\documentclass[twocolumn,aps,superscriptaddress,showpacs,floatfix,prc]{revtex4}
\usepackage{mathrsfs}
\usepackage{amssymb}
\usepackage{amsmath}
\usepackage{graphicx}
\usepackage[normalem]{ulem}
\usepackage[dvips]{color}
\usepackage{bm}
\usepackage{longtable}

\renewcommand\sout{\bgroup \color{red} \ULdepth=-.5ex \ULset}

\renewcommand{\v}[1]{\textbf{#1}}
\renewcommand{\rm}[1]{\textrm{#1}}
\renewcommand{\d}{\mathrm{d}}

\begin{document}

\title{Nucleon Effective E-Mass in Neutron-Rich Matter from the Migdal-Luttinger Jump}

\author{Bao-Jun Cai\footnote{Email:landau1908feynman1918@gmail.com}}
\affiliation{Department of Physics and Astronomy, Texas A$\&$M
University-Commerce, Commerce, TX 75429-3011, USA}
\author{Bao-An Li\footnote{Corresponding author: Bao-An.Li$@$tamuc.edu}}
\affiliation{Department of Physics and Astronomy, Texas A$\&$M
University-Commerce, Commerce, TX 75429-3011, USA}
\date{\today}

\begin{abstract}
The well-known Migdal-Luttinger theorem states that the jump of the
single-nucleon momentum distribution at the Fermi surface is equal to the
inverse of the nucleon effective E-mass. Recent
experiments studying short-range correlations (SRC) in nuclei using
electron-nucleus scatterings at the Jefferson National Laboratory (JLAB)
together with model calculations constrained significantly the
Migdal-Luttinger jump at saturation density of nuclear matter. We
show that the corresponding nucleon effective E-mass is consequently constrained to
$M_0^{\ast,\rm{E}}/M\approx2.22\pm0.35$ in symmetric nuclear matter
(SNM) and the E-mass of neutrons is smaller than that of protons in
neutron-rich matter. Moreover, the average depletion of the nucleon
Fermi sea increases (decreases)  approximately linearly with the
isospin asymmetry $\delta$ according to
$\kappa_{\rm{p}/\rm{n}}\approx 0.21\pm0.06 \pm (0.19\pm0.08)\delta$
for protons (neutrons). These results will help improve our knowledge
about the space-time non-locality of the single-nucleon potential in
neutron-rich nucleonic matter useful in both nuclear physics and astrophysics.
\end{abstract}

\pacs{21.65.Ef, 24.10.Ht, 21.65.Cd} \maketitle

\section{Introduction}
In the framework of Landau Fermi liquid
theory\,\cite{Lan57,Abr59,Abr63,Noz64,Lif80,Bay91,Pin94,Neg98}, the
(Landau) effective mass of a Fermion is a fundamental quantity
describing to leading order effects related to the space-time
nonlocality of the underlying interactions and the Pauli exclusion
principle. The study of nucleon effective mass in finite nuclei
and/or infinite nuclear matter has a long history because of the
great challenges involved and its significance for both nuclear
physics and astrophysics, see, e.g., refs.\,\cite{Jeu76,Mah85,jamo}
for earlier reviews. Moreover, there are interesting new issues related
to the isospin dependence of space-time nonlocality determining,
such as the neutron-proton effective mass splitting and their
interaction cross sections in neutron-rich nucleonic matter, see,
e.g., ref.\,\cite{LiBA15} for a recent review. Despite of the
impressive progress made in this field, our current knowledge on the
nucleon effective mass especially its isospin dependence is still
rather poor. It is thus widely recognized that better knowledge on the nucleon effective mass is critical for
us to make further progress in solving many other interesting
problems in both nuclear physics and astrophysics. For example, the
isospin dependence of space-time non-locality affects the symmetry
energy of asymmetric nuclear matter
(ANM)\,\cite{LiBA98,Dan02,Bar05,Ste05,Che07a,LCK08,Tsa12,Che14,Gle00,Lat04,Lat12,Lat14,LiBA13,EPJA},
the momentum dependence of both the isoscalar and isovector parts of the single-nucleon potential \cite{Bom01,Li04,LiBA04,zuo05,Riz05,Gio10,Feng12,Zhang14,Xie14} and the in-medium nucleon-nucleon scattering cross sections \cite{Neg81,Pan91,Gale,Li&Chen05} used in simulating heavy-ion collisions especially
those induced by rare isotopes, the level densities and thermal properties of hot nuclei \cite{Pra83,Shl91,Bob05,Beh11,Bob14,Jun15} as well as the cooling rate and
transport properties of neutron stars\,\cite{Hae07,HFZhang}.

The effective mass of a nucleon $J=(\rm{n,p})$ can be calculated
from the derivative of its potential $U_J$ with respect to either
its energy $E$ or momentum $\v{k}$\,\cite{jamo}
\begin{eqnarray}\label{em1}
\frac{M^{*}_{J}}{M}&=&1-\frac{\d U_{J}(\v{k}(E),E,\rho,\delta)}{\d
E}\\\nonumber &=&\left[1+\frac{M}{\hbar^2k_{\rm{F}}^J}\frac{\d
U_{J}(\v{k},E(\v{k}),\rho,\delta)}{\d|\v{k}|}\right]^{-1}
\end{eqnarray}
where $M$ is the average mass of nucleons in free-space. Moreover,
we take $|\v{k}|=k_{\rm{F}}^J$ in this work with
$k_{\rm{F}}^J=(1+\tau_3^J\delta)^{1/3}\cdot k_{\rm{F}}$ and
$k_{\rm{F}}=(3\pi^2\rho/2)^{1/3}$ being the nucleon Fermi momentum
in symmetric nuclear matter at density $\rho$, $\tau_3^{J}=+1$ or
$-1$ for neutrons or protons and
$\delta=(\rho_{\rm{n}}-\rho_{\rm{p}})/(\rho_{\rm{n}}+\rho_{\rm{p}})$
is the isospin asymmetry of the medium. The ${M^{*}_{J}}/{M}$ at
saturation density $\rho_0$ can be extracted from the
energy/momentum dependence of nucleon optical potentials using an
on-shell energy-momentum dispersion
relation\,\cite{ZHLI,LiX13,LiX15}. It is known  that the total
nucleon effective mass can be decomposed
into\,\cite{Jeu76,Mah85,jamo}
\begin{equation}\label{Tmass}
\frac{M^{\ast}_{J}}{M}=\frac{M^{\ast,\rm{E}}_{J}}{M}\cdot\frac{M^{\ast,\rm{k}}_{J}}{M}
\end{equation}
with
\begin{equation}
\frac{M^{\ast,\rm{E}}_{J}}{M}=1-\frac{\partial U_{J}}{\partial
E}~\rm{and}~
\frac{M^{\ast,\rm{k}}_{J}}{M}=\left[1+\frac{M_{J}}{|\v{k}|}\frac{\partial
U_{J}}{\partial|\v{k}|}\right]^{-1}
\end{equation}
the nucleon E-effective mass (E-mass) and k-effective mass (k-mass) to characterize the energy and momentum dependence of the single-nucleon potential $U_J$ due to the
non-locality in time and space of the underlying interaction, respectively.
\begin{figure}[h!]
\centering
  \includegraphics[width=8.cm]{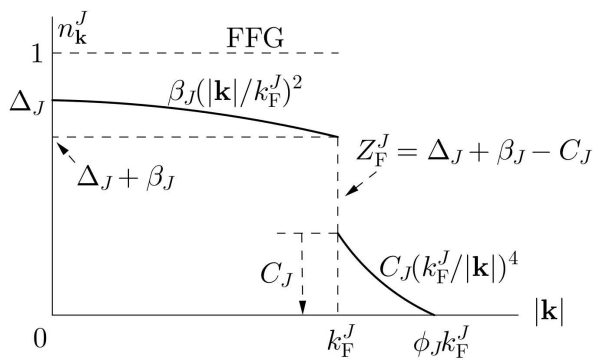}
  \caption{A sketch of the single nucleon momentum distribution with a high momentum tail used in this work.}
  \label{mom-dis}
\end{figure}

Most experiments and phenomenological models probe only the total
effective mass $M^{\ast}_{J}/M$
\cite{LiBA15,Jeu76,Mah85,jamo,ZHLI,LiX13,LiX15,Zha15c}. From Eq.\
(\ref{Tmass}), it is seen that an independent determination of
either the E-mass or k-mass together with the total effective mass
will then allow us to know all three kinds of nucleon effective masses.
Interestingly, the Migdal-Luttinger
theorem\,\cite{Mig57,Lut60} connects the nucleon E-mass directly with
the jump (discontinuity) $Z_{\rm{F}}^J\equiv
n_{\v{k}}^J(k^J_{\rm{F}-0})-n_{\v{k}}^J(k^J_{\rm{F}+0})$ of the
single-nucleon momentum distribution $n_{\v{k}}^J$ at the Fermi
momentum $k^J_{\rm{F}}$ illustrated in Fig. \ref{mom-dis} via
\begin{equation}\label{ML}
{M_{J}^{\ast,\rm{E}}}/{M}=1/Z_{\rm{F}}^J.
\end{equation}
The nuclear physics community has devoted much efforts to probing
the depletion of the nucleon Fermi sphere by using transfer, pickup
and (e,e$'$p) reactions. Results of these studies normally given in
terms of the nucleon spectroscopic factors can constrain the
$n_{\v{k}}^J(k^J_{\rm{F}-0})$\,\cite{Jeu76,Mah85,jamo}. On the other
hand, quantitative information about both the shape and magnitude of
the high-momentum tail (HMT) above the Fermi surface have been
extracted recently from analyzing cross sections of both inclusive
and exclusive electron-nucleus
scatterings\,\cite{Hen15,Hen14,Col15,Egi06,Pia06} as well as
medium-energy photonuclear absorptions\,\cite{Wei15,Wei15a},
providing a constraint on the $n_{\v{k}}^J(k^J_{\rm{F}+0})$. These
experimental results together with model analyses provide a
significant empirical constraint on the Migdal-Luttinger jump. In
this work, we show that the corresponding nucleon E-mass is consequently
constrained to $M^{\ast,\rm{E}}/M\approx2.22\pm0.35$ in symmetric
nuclear matter and the E-mass of neutrons is smaller than that of
protons in neutron-rich matter. Moreover, the average depletion of
the nucleon Fermi sea increases (decreases) approximately linearly
with the isospin asymmetry $\delta$ according to
$\kappa_{\rm{p}/\rm{n}}\approx 0.21\pm0.06 \pm (0.19\pm0.08)\delta$
for protons (neutrons).

 \section{The Single-Nucleon Momentum Distribution Function in cold neutron-rich nucleonic matter}
We briefly recall here the main features of $n_{\v{k}}^J$ used in
the present work\,\cite{Cai15}. It is well known that the SRC due to
tensor components and/or the repulsive core of nuclear forces leads
to a high (low) momentum tail (depletion) in the single-nucleon
momentum distribution above (below) the nucleon Fermi momentum in
cold nucleonic matter, see refs.\,\cite{Bethe,Ant88,Arr12,Cio15} for
comprehensive reviews. It has been found from analyzing
electron-nucleus scattering data that the percentage of nucleons in
the HMT is about 25\% in SNM but decreases gradually to about only
1\% in pure neutron matter (PNM)\,\cite{Hen14,Hen15}.

We parameterize the single-nucleon momentum distribution
in cold ANM with
\begin{equation}\label{MDGen}
n^J_{\v{k}}(\rho,\delta)=\left\{\begin{array}{ll}
\Delta_J+\beta_J{I}\left(\displaystyle{|\v{k}|}/{k_{\rm{F}}^J}\right),~~&0<|\v{k}|<k_{\rm{F}}^J,\\
&\\
\displaystyle{C}_J\left({k_{\rm{F}}^{J}}/{|\v{k}|}\right)^4,~~&k_{\rm{F}}^J<|\v{k}|<\phi_Jk_{\rm{F}}^J.
\end{array}\right.
\end{equation}
As sketched in Fig. \ref{mom-dis}, the $\Delta_J$ measures the
depletion of the Fermi sphere at zero momentum with respect to the
free Fermi gas (FFG) model prediction while the $\beta_J$ is the
strength of the momentum dependence $I(\v{k}/k_{\rm{F}}^J)$ of the
depletion near the Fermi surface. The parameters $\Delta_J$, $C_J$,
$\phi_J$ and $\beta_J$ depend linearly on the isospin asymmetry
according to $Y_J=Y_0(1+Y_1\tau_3^J\delta)$\,\cite{Cai15}. The
amplitude ${C}_J$ and the cutoff coefficient $\phi_J$ determine the
fraction of nucleons in the HMT via
\begin{equation}\label{xPNM}
x_J^{\rm{HMT}}=3C_{{J}}\left(1-\frac{1}{\phi_{{J}}}\right).
\end{equation}
The normalization condition $
[{2}/{(2\pi)^3}]\int_0^{\infty}n^J_{\v{k}}(\rho,\delta)\d\v{k}=\rho_J={(k_{\rm{F}}^{J})^3}/{3\pi^2}
$ requires that only three of the four parameters, i.e., ${C}_J$,
$\phi_J$, $\beta_J$ and $\Delta_J$, are independent. Here we choose
the first three as independent and determine the $\Delta_J$
by\,\cite{Cai15}
\begin{equation}\label{DeltaJ}
\Delta_J=1-\frac{3\beta_J}{(k_{\rm{F}}^J)^3}\int_0^{k_{\rm{F}}^J}I\left(\frac{k}{k_{\rm{F}}^J}\right)k^2\d
k-3{C}_J\left(1-\frac{1}{\phi_J}\right).
\end{equation}

The ${C}/{|\mathbf{k}|^4}$ shape of the HMT both for SNM and PNM is
strongly supported by recent findings theoretically and
experimentally. Combining results of analyzing the
$\rm{d}(\rm{e}, \rm{e}^{\prime}\rm{p})$ cross sections \cite{Hen15} with an evaluation of medium-energy
photonuclear absorption cross sections\,\cite{Wei15} leads to a
value of $C_0\approx0.161\pm0.015$. With this $C_0$ and the value of
$x^{\rm{HMT}}_{\rm{SNM}}=28\%\pm4\%$\,\cite{Hen14,Hen15,Hen15b}
obtained from systematic analyses of inclusive (e,e$'$) reactions
and data from exclusive two-nucleon knockout reactions, the HMT
cutoff parameter in SNM is determined to be
$\phi_0=(1-x_{\rm{SNM}}^{\rm{HMT}}/3{C}_0)^{-1}=2.38\pm0.56$\,\cite{Cai15}.
The value of $C_{\rm{n}}^{\rm{PNM}}=C_0(1+C_1)$ is extracted by
applying the adiabatic sweep theorem\,\cite{Tan08} to the EOS of PNM
predicted by microscopic many-body
theories\,\cite{Sch05,Epe09a,Tew13,Gez13,Gez10} as well as that from
the EOS of Fermi systems under unitary
condition\,\cite{Tan08,Stew10}. More quantitatively, we obtained
$C_{\rm{n}}^{\rm{PNM}}\approx0.12$ and subsequently
$C_1=-0.25\pm0.07$\,\cite{Cai15}. By inserting the value of
$x_{\rm{PNM}}^{\rm{HMT}}=1.5\%\pm0.5\%$\,\cite{Hen14,Hen15,Hen15b}
and the $C_{\rm{n}}^{\rm{PNM}}$ into Eq. ({\ref{xPNM}), the high momentum
cutoff parameter for PNM is determined to be
$\phi_{\rm{n}}^{\rm{PNM}}\equiv
\phi_0(1+\phi_1)=(1-x_{\rm{PNM}}^{\rm{HMT}}/3C_{\rm{n}}^{\rm{PNM}})^{-1}=1.04\pm0.02$\,\cite{Cai15}.
Consequently, we get $\phi_1=-0.56\pm0.10$\,\cite{Cai15} by using
the $\phi_0$ determined earlier. Moreover, a quadratic momentum-dependence
$I(k/k_{\rm{F}}^J)=(k/k_{\rm{F}}^J)^2$ is adopted\,\cite{Cai15}
from predictions of some nuclear many-body theories\,\cite{Cio96}, then Eq.
(\ref{DeltaJ}) gives us $\Delta_J = 1-3\beta_J/5 - 3C_J (1 -
1/\phi_J )$. Specifically, we have $\beta_0 = (5/3)[1 -\Delta_0 -
3C_0(1 - \phi_0^{-1})]=(5/3)[1-\Delta_0- x^{\rm{HMT}}_{\rm{SNM}}]$
for SNM. Then, using the predicted value of $\Delta_0 \approx
0.88\pm0.03$\,\cite{Pan97,Yin13,Fan84} and the experimental value of
$x^{\rm{HMT}}_{\rm{SNM}} \approx 0.28 \pm 0.04$, the value of
$\beta_0$ is estimated to be about $-0.27\pm0.08$. Similarly, the
condition $\beta_J = \beta_0(1 + \beta_1\tau_3^J\delta) < 0$, i.e.,
$n^J_{\v{k}}$ is a decreasing function of momentum towards
$k_{\rm{F}}^J$, indicates generally that $|\beta_1|\leq1$. For more details of
these parameters, see ref.\,\cite{Cai15}.

\begin{figure}[h!]
\centering
  \includegraphics[width=8.cm]{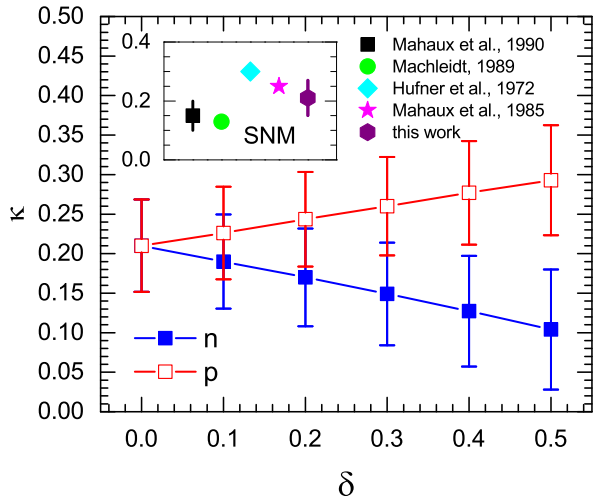}
  \caption{(Color Online) The average depletion of the neutron and proton Fermi surface as a function of isospin asymmetry in neutron-rich matter.
  The inset shows the average nucleon depletion in symmetric matter from this work and earlier studies\,\cite{Mah90,Mac89,Huf72,Mah85}.}
  \label{Fig-kappa}
\end{figure}
The average depletion of the Fermi sphere in asymmetric nuclear
matter
\begin{align}
\kappa_J=&1-\Delta_J-\frac{1}{k_{\rm{F}}^J}\int_0^{k_{\rm{F}}^J}\beta_J\left(\frac{|\v{k}|}{k_{\rm{F}}^J}\right)^2\d
k\notag\\
=&\frac{4}{15}\beta_J+3C_J\left(1-\frac{1}{\phi_J}\right)
\end{align}
depends strongly on the tensor part of the nucleon-nucleon
interaction\,\cite{Fan84,Ram90}. It provides a quantitative measure
of the validity of the Hugenholtz-Van Hove (HVH) theorem\,\cite{HVH}
and more generally independent particle models. A deeper depletion
indicates a more serious violation of the HVH
theorem\,\cite{Mah90,Jon91,Mac89,Huf72}. Experimentally, it can be
measured by using the nucleon spectroscopic factor from transfer,
pickup and (e,e$'$p) reactions\,\cite{Mah90}. A well-known example
is the finding that mean-field models overpredict the occupation of
low-momentum nucleon orbitals compared to data of electron
scatterings on nuclei from $^{7}$Li to $^{208}$Pb by about 30-40\%
due to the neglect of correlations\,\cite{Lap93}. The $\kappa_J$ is
also believed to determine the rate of convergence of the hole-line
expansion of the nuclear potential\,\cite{Bethe,Mah85,Huf72,Mah90}.
In Fig. \ref{Fig-kappa}, the average depletion of the neutron and
proton Fermi surface is shown separately as a function of isospin
asymmetry in neutron-rich matter. It is interesting to see that the
neutron/proton depletion decreases/increases with $\delta$
approximately linearly, indicating that protons with energies near the Fermi
surface experience larger correlations with increasing asymmetry
in qualitative agreement with findings from both analyses using microscopic
many-body theories\,\cite{Ram90,Yin13} and phenomenological
models\,\cite{Sar14}. This is also consistent with experimental findings from earlier studies of nucleon spectroscopic factors \cite{Gade},
dispersive optical model analyses of proton-nucleus scatterings \cite{Bob06} and the neutron-proton dominance model analyses of electron-nucleus scattering experiments \cite{Hen14}.
More quantitatively, the neutron-proton
splitting of the $\kappa_J$ is approximately
$\kappa_\rm{n}-\kappa_\rm{p}\approx[8\beta_0\beta_1/15+6C_0\phi_1/\phi_0+6C_0C_1(1-\phi_0^{-1})]\delta\approx(-0.37\pm0.16)\delta$.
For symmetric nuclear matter, we have
$\kappa=4\beta_0/15+x_{\rm{SNM}}^{\rm{HMT}}\approx0.21\pm0.06$
comparable with the results obtained earlier from other
studies \,\cite{Mah90,Mac89,Huf72,Mah85}, as shown in the inset of
Fig. \ref{Fig-kappa}.

Several consequences of the SRC modified nucleon momentum
distribution have been studied recently. In particular, it was found
that the nucleon kinetic symmetry energy is reduced compared to the
FFG model prediction\,\cite{CXu11,CXu13,Vid11,Lov11,Car12,Rio14,Car14,Hen15b,Cai15,Cai15c}.
This has important consequences on isovector observables in
heavy-ion collisions\,\cite{Hen15b,Li15,Yon15,Li15a} and on the
critical densities for forming different charge states of
$\Delta(1232)$ resonances in neutron stars\,\cite{Cai15a,Dra14}.
Moreover, the SRC was also found to enhance the isospin-quartic term
in the kinetic energy of ANM within both
non-relativistic\,\cite{Cai15} and relativistic
models\,\cite{Cai15c}. Very recently, it was shown that the SRC-induced depletion of the nucleon
Fermi surface affects significantly the neutrino emissivity, heath capacity and neutron superfluidity
in neutron stars \cite{Dong15}.

Before going further, it is necessary to address a possible drawback of our parameterization for the single-nucleon momentum distribution in Eq. \ref{MDGen}.
To our best knowledge, the original derivations \cite{Mig57,Lut60} of the Migdal-Luttinger theorem do not explicitly require the derivative of the momentum disturibution
 $[\d n_{\v{k}}/\d k]_{k=k_{\rm{F}}\pm0}$ at the Fermi momentum $k_{\rm{F}}$ to be $-\infty$.
 Of course, one can associate mathematically loosely the finite drop in $n_{\v{k}}$
over zero increase in momentum at $k_{\rm{F}}$ to a slope of
$-\infty$. Some later derivations using various approximation
schemes, such as the ``derivative expansion" in which the momentum
distribution is expressed in terms of energy derivatives of the mass
operator by Mahaux and Saror \cite{Mah92}, have shown that the slope
of the momentum distribution should have the asymptotic behavior of
$[\d n_{\v{k}}/\d k]_{k=k_{\rm{F}}\pm0}=-\infty$. Similar to many
other calculations including some examples given in refs.
\cite{Mah85,Mah92}, our parameterization of Eq. \ref{MDGen} does not
have such behavior from neither side of the discontinuity. However,
similar to what has been done in ref. \cite{Bal90}, in
parameterizing the $n_{\v{k}}$ both above and below the $k_{\rm{F}}$
one can add a term that is vanishingly small in magnitude but
asymptotically singular in slope at $k_{\rm{F}}$, such as
$\eta\cdot(\frac{k-k_{\rm{F}}}{\Lambda})\cdot
\ln(\frac{k-k_{\rm{F}}}{\Lambda})$ where $\eta$ is a constant much
smaller than $\Delta_0$ and $C_0$. Of course, one then has to
determine the totally 4 additional parameters ($\eta$ and $\Lambda$
for neutrons and protons above and below their  respective Fermi
momenta) and readjust the other parameters already used in Eq.
\ref{MDGen}. This has the potential of reducing the error bars of
the quantities we extract but requires more experimental
information. Unfortunately, the analyses of existing experimental
data we mentioned above have so far not considered such corrections.
While we do not expect the corrections will affect significantly the
size of the Migdal-Luttinger jump since they have vanishingly small
magnitudes at $k_{\rm{F}}$, our description about the discontinuity
of $n_{\v{k}}$ at $k_{\rm{F}}$ certainly should be investigated
further and possibly improved in the future. For the present
exploratory study using information from phenomenological model
analyses of limited experimental data, we feel that the
parameterization of Eq. \ref{MDGen} is good enough.

\section{Nucleon E-effective Mass and Its Isospin Splitting in neutron-rich nucleonic matter}
We now turn to the nucleon E-mass obtained through the
Migdal-Luttinger theorem of Eq. (\ref{ML}). In terms of the
parameters describing the single-nucleon momentum distribution
$n_{\v{k}}^J$, we have
$Z_{\rm{F}}^J=\Delta_J+\beta_J-C_J=1+2\beta_J/5-4C_J+3C_J\phi_J^{-1}$.
For SNM, it is given by
$Z_{\rm{F}}^0=1+2\beta_0/5-4C_0+3C_0\phi_0^{-1}=1+2\beta_0/5-C_0-x_{\rm{SNM}}^{\rm{HMT}}$,
then using the values for $\beta_0$, $\phi_0$, $C_0$ and
$x_{\rm{SNM}}^{\rm{HMT}}$ given above, we obtain a value of
$M_0^{\ast,\rm{E}}/M\approx2.22\pm0.35$. Shown in Fig.
\ref{Fig-Emass0} with the filled squares are the extracted nucleon
E-mass in SNM within the uncertain range of the $\beta_0$ parameter. It is
seen that the variation of $M_0^{\ast,\rm{E}}/M$ with $\beta_0$ is
rather small. For comparisons, also shown are earlier predictions
based on (1) a semi-realistic parametrization through a relative
s-wave exponential nucleon-nucleon interaction potential (red dash line)\,\cite{Ber81}, (2) a Green's function method considering collective effects due to the coupling
of nucleons with the low-lying particle-hole excitations of the
medium (green solid line)\,\cite{Bla81}, (3) a correlated
basis function (CBF) method using the Reid and Bethe-Johnson
potentials} (black and magenta solid lines)\,\cite{Kro81,Jac82}, (4)
two non-relativistic models with the Paris nuclear potential (purple and
red solid line)\,\cite{Gra87,Bal90}, (5) a low density expansion of
the optical potential (orange solid line)\,\cite{Sar77} and (6) a
relativistic Dirac-Brueckner approach (dash black
line)\,\cite{Jon91}. While we are unable to comment on possible
origins of the different model predictions and their differences
from the empirical values presented here, to our best knowledge, it
is the first time that the nucleon E-mass is extracted using the Migdal-Luttinger theorem from the single-nucleon momentum distribution
constrained by experiments phenomenologically.
\begin{figure}[h!]
\centering
  \includegraphics[width=8.5cm]{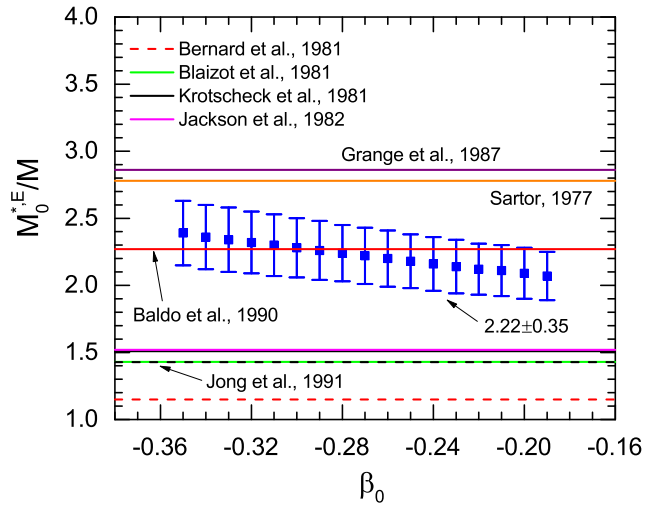}
  \caption{(Color Online) The nucleon effective E-mass in symmetric nuclear matter (blue lines with error bars) at normal density
  within the uncertainty range of the shape parameter $\beta_0$ of the nucleon momentum distribution extracted using the Migdal-Luttinger theorem in this work
  in comparison with predictions of earlier studies\,\cite{Ber81,Bla81,Kro81,Jac82,Gra87,Bal90,Sar77,Jon91}, see detailed descriptions in the text.}
  \label{Fig-Emass0}
\end{figure}

\begin{figure}[h!]
\centering
  \includegraphics[width=8.cm]{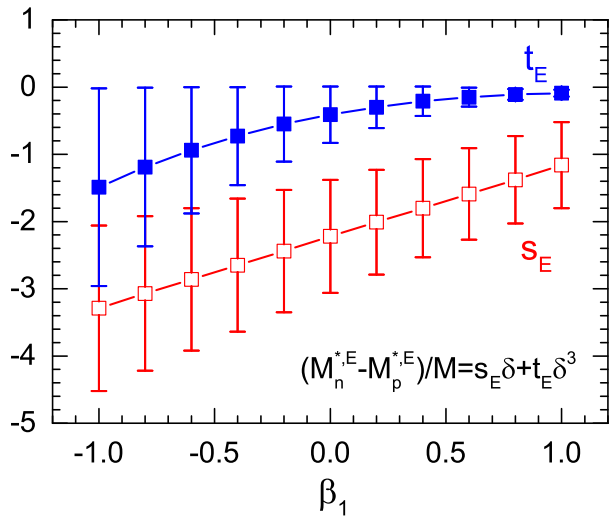}
  \caption{(Color Online) The linear and cubic splitting functions $s_{\rm{E}}$ and $t_{\rm{E}}$ at normal density within the uncertain range of the $\beta_1$-parameter characterizing the isospin-dependence of the nucleon momentum distribution near the Fermi surface.}
  \label{Fig-Splitting}
\end{figure}

In neutron-rich nucleonic matter, an interesting quantity is the neutron-proton E-mass splitting generally expressed as
\begin{equation}
\frac{M_{\rm{n}}^{\ast,\rm{E}}-M_{\rm{p}}^{\ast,\rm{E}}}{M}=s_{\rm{E}}\delta+t_{\rm{E}}\delta^3
+\mathcal{O}(\delta^5)
\end{equation}
where $s_{\rm{E}}$ and $t_{\rm{E}}$ are the linear and cubic
splitting functions, respectively. The latter generally depend on the nucleon momentum and the density of the medium.
Shown in Fig. \ref{Fig-Splitting} are the values of $s_{\rm{E}}$ and $t_{\rm{E}}$ at the nucleon Fermi momentum in nuclear matter at $\rho_0$ within the uncertainty range of the $\beta_1$-parameter. More quantitatively, at the lower limit, mid-value and upper limit of the $\beta_1$-parameter, we have $s_{\rm{E}}(\beta_1=-1)\approx-3.29\pm1.23$,
$t_{\rm{E}}(\beta_1=-1) \approx-1.49\pm1.47$,
$s_{\rm{E}}(\beta_1=0)\approx-2.22\pm0.84$, $t_{\rm{E}}(\beta_1=0)\approx-0.41\pm0.42$,
$s_{\rm{E}}(\beta_1=1)\approx-1.16\pm0.64$ and
$t_{\rm{E}}(\beta_1=1)\approx-0.09\pm0.05$, respectively. We note that the cubic splitting function
$t_{\rm{E}}$ generally can not be neglected (e.g., for $\beta_1=0$,
$t_{\rm{E}}/s_{\rm{E}}\approx18\%$), and it may have sizable effects
on the cooling and thermal properties of neutron stars\,\cite{Hae07,Dong15}.

An important feature shown in Fig. \ref{Fig-Splitting} is that in
neutron-rich nucleonic matter, the E-mass of a neutron is
smaller than that of a proton, i.e.,
$M_{\rm{n}}^{\ast,\rm{E}}<M_{\rm{p}}^{\ast,\rm{E}}$. However, the neutron-proton E-mass splitting in ANM has an appreciable dependence on the
largely uncertain $\beta_1$-parameter characterizing the isospin-dependence of the nucleon momentum distribution near the Fermi surface.
Unfortunately, currently there exists no reliable constraint on the
parameter $\beta_1$. Thus, it is interesting to mention that there are experimental efforts to
measure the isospin dependence of the nucleon spectroscopic factors
using direct reactions with radioactive beams\,\cite{Jenny} and
the isospin-dependence of SRC with both electrons and
hadrons\,\cite{SRC-Coms}. These experiments have the potential to
constrain the $\beta_1$ and thus the neutron-proton E-mass splitting
in neutron-rich matter.

\section{Summary}
In summary, using the Migdal-Luttinger theorem relating the discontinuity of the
single-nucleon momentum distribution function at the Fermi surface with the nucleon E-mass,
we have extracted the latter and its isospin splitting in neutron-rich nucleonic matter at normal density using the single-nucleon momentum distribution constrained by recent experiments at the JLAB. We found that the nucleon E-mass in SNM is $M_0^{\ast,\rm{E}}/M\approx2.22\pm0.35$ while in neutron-rich matter the E-mass of neutrons is smaller than that of protons. Moreover, the average depletion of the nucleon Fermi sea increases (decreases) approximately linearly with the isospin asymmetry $\delta$ according to $\kappa_{\rm{p}/\rm{n}}\approx 0.21\pm0.06 \pm (0.19\pm0.08)\delta$ for protons (neutrons). These results provide useful references for microscopic nuclear many-body theories and will help improve our knowledge about the space-time non-locality of the single-nucleon potential in neutron-rich nucleonic matter.

\section*{Acknowledgement}
We would like to thank Isaac Vida\~na and William G. Newton for helpful discussions. This work was supported in part by the U.S. National Science Foundation under Grant No. PHY-1068022, the U.S. Department of Energy's Office of Science under Award Number DE-SC0013702 and the National Natural Science Foundation of China under grant no. 11320101004.

\end{document}